\documentclass[useAMS,usenatbib,onecolumn]{mn2e}

\usepackage{mathrsfs}
\usepackage{amsmath}
\usepackage{graphicx}

\newcommand{\D}{\mathrm{D}}
\newcommand{\curl}{\mathrm{curl}}
\newcommand{\df}{\lambda}
\newcommand{\ed}{\mathscr{E}}

\begin{document}

\title[The general relativistic MHD dynamo equation]{The general relativistic MHD dynamo
  equation}

\author[M.\ Marklund and C.\ Clarkson]{M.\ Marklund$^{1, 2}$ and C.\ A.\ Clarkson$^3$ \\
{$^1\,$Department of Physics, Ume{\aa} University,
  SE--901 87 Ume{\aa}, Sweden} \\
{$^2\,$Centre for Fundamental Physics,
  Rutherford Appleton Laboratory,
  Chilton Didcot, Oxfordshire, OX11 0QX, UK} \\
{$^3\,$Institute of Cosmology and Gravitation, University of
Portsmouth, Portsmouth, PO1 2EG, UK} \\
\rm Email: \texttt{mattias.marklund@physics.umu.se} and
\texttt{chris.clarkson@port.ac.uk}}

\date{\today}

\pagerange{\pageref{firstpage}--\pageref{lastpage}} \pubyear{2004}

\maketitle

\begin{abstract}
    The magnetohydrodynamic dynamo equation is derived within general
    relativity, using the covariant $1+3$ approach, for a plasma with finite
    electric conductivity. This formalism allows for a clear division and
    interpretation of plasma and gravitational effects, and we have not restricted
    to a particular spacetime geometry. The results should be of interest in
    astrophysics and cosmology, and the formulation is well suited
    to gauge invariant perturbation theory. Moreover, the dynamo equation is
    presented in some specific limits. In particular, we consider the
    interaction of gravitational waves with magnetic fields, and present
    results for the evolution of the linearly growing electromagnetic
    induction field, as well as the diffusive damping of these fields.
\end{abstract}


\section{Introduction}

Magnetic fields play a vital role in many astrophysical systems
and in cosmology, and the possible sources for the fields on
different scales has been the focus of immense research over the
years \citep{kron,han,park,zel,beck}. In particular, the origin of
large scale cosmological fields is still not clearly understood.
Although there are well understood mechanisms for amplifying a
given seed field \citep{gras,wid}, the explanation of the seed
sources is more cumbersome, especially in the case of an
inflationary scenario. A wide range of possible physical
explanations for establishing a seed field of high enough field
strength has been proposed in the literature (see, e.g.,
\citet{har,hog,vach,gas,sigl,joy,dav,cal,dim,bet,mmnr}). The main
mechanism behind the amplification of the seed fields is the
dynamo \citep{park,zel,beck}, for which the fluid motion manages
to amplify even very weak fields on very short time scales. The
dynamo mechanism derives from the magnetohydrodynamic (MHD)
approximation, and it has been analysed vigorously within, e.g.,
expanding cosmological models (for a selection, see
\citet{fen,sil,gai,jed,bra,sub} and references therein).

Cosmic magnetic fields are observed on all but the largest scales.
On the other hand, gravitational waves have yet to be directly
detected. Although strong evidence exists in their favour (such as
the PSR 1913+16 binary pulsar observations \citep{hul}), the very
weak interaction of gravitons with light and matter makes an earth
based detection difficult. Currently, gravitational wave
detectors, both of interferometer and bar type, are coming
on-line, and are expected to start collecting important
astronomical data, opening up a new window for observational
astrophysics and cosmology. As gravitational wave information is
gathered, the detailed study of typical wave signatures will be of
major importance. Thus, knowledge of both direct gravitational
wave signature, and indirect signatures, e.g., electromagnetic
waves induced by gravity waves~\citep{ringdown}, will facilitate
the study of gravitational waves, and is likely to bring new
understanding to the observations.

The interaction of gravitational waves, and other general
relativistic gravity effects, with electromagnetic fields and
plasmas have attracted a broad interest over the last few years,
because its possible application to both astrophysics and
cosmology (e.g., \citet{bla,tho,mac,mar,tsa,ringdown,tom,%
tsa-bar,tsa-maa,tsa01}). Here, we
will derive the general form of the dynamo equation, using the
$1+3$ covariant formalism \citep{EvE}, and analyse its coupling to
matter and spacetime geometry. This formalism allows for a clear
cut interpretation of gravitational vs.\ purely MHD effects, and
also presents the influence of the gravitational kinematics on the
MHD modes for straightforward comparison. Specifically, we will
consider the propagation of gravitational waves through a
resistive magnetohydrodynamic plasma, and derive covariant
equations for the propagation of a resulting magnetohydrodynamic
wave. Here, a word of caution concerning the use of the phrase `dynamo'
is in place. While speaking of the dynamo equation, a general
evolution equation for the magnetic field in diffusive media is
assumed, while the dynamo mechanism is a particular application
of this general equation. Indeed, since the dynamo mechanism is
without question the most prominent feature of the general
dynamo equation, we would like to stress that although we
derive the general dynamo equation, we will not apply this
to the dynamo mechanism. Instead, we will investigate
what could appropriately be termed induction fields.
The waves presented here represents linearly growing induction fields,
and are thus not due to the proper dynamo mechanism which
instead gives rise to rapid amplification of seed magnetic field due to the
vortical fluid motion. Furthermore, the effects of a finite conductivity
is presented.
The equations are analysed numerically, and we discuss
possible applications to astrophysical and cosmological systems.

\section{Basic equations}

When analysing low frequency magnetised plasma phenomena, MHD
gives an accurate and computationally economical description.
Specifically, a simple plasma model is obtained if the
characteristic MHD time scale is much longer than both the plasma
oscillation and plasma particle collision time scales, and the
characteristic MHD length scale is much longer than the plasma
Debye length and the gyro radius. These assumptions will make it
possible to describe a two-component plasma in terms of a
one-fluid description, as the electrons can be considered
inertialess (see also, e.g., \citet{ser,mor,mor2,pap,pap2}).  The
one-fluid description means a tremendous computational
simplification, especially for complicated geometries. Moreover,
if the mean fluid velocity, the mean particle velocity, and the
Alfv\'en velocity is much smaller than the velocity of light in
vacuum, the description becomes non-relativistic and simplifies
further.

\subsection{Covariant theory}

The 1+3 covariant approach relies on the introduction of a
timelike vector field $u^a$: $u_au^a=-1$, with which all tensorial
objects are split into their invariant `time' and `space' parts:
scalars, 3-vectors, and projected, symmetric, trace-free (PSTF)
tensors (see~\citet{EvE} for a comprehensive review). This allows
us to write the coupled Einstein-Maxwell equations in a relatively
simple and intuitive fashion. Here, we introduce consistent
3-vector notation of Euclidean vector calculus, making the fully
general relativistic equations relatively easy to read.

In an arbitrary curved spacetime, in units such that $c = 1$ and
$G = 1/8\pi$, Maxwell's equations take the form
\citep{tsa-bar,EvE,mar2}
\begin{eqnarray}
     \dot{B}^{\langle a\rangle}+ {\curl}\,E^{a} &=& -\left( \tfrac{2}{3}\Theta h^{ab} -
    \sigma^{ab} + \epsilon^{abc}\omega_{c} \right) B_{b} -
    \epsilon^{abc}\dot{u}_{b}E_{c}  , \label{eq:m1} \\
     -\dot{E}^{\langle a\rangle} + {\curl}\,B^{a} &=& \mu_0j^{\langle a\rangle} +
    \left( \tfrac{2}{3}\Theta h^{ab} - \sigma^{ab} +
    \epsilon^{abc}\omega_{c} \right) E_{b} -
    \epsilon^{abc}\dot{u}_{b}B_{
    c} , \label{eq:m2} \\
    \D_{a}E^{a} &=& 2\omega_{a}B^{a} , \label{eq:m3} \\
    \D_{a}B^{a} &=& -2\omega_{a}E^{a} , \label{eq:m4}
\end{eqnarray}
where $\curl\,X^{a} \equiv \epsilon^{abc}\D_{b}X_{c}$ and
$\dot{X}^{\langle a\rangle} \equiv h^{ab}\dot{X}_b$ for any
spatial vector $X^{a}$, and $h_{ab} = g_{ab} + u_au_b$ is the metric on
the local rest space orthogonal to the observer four-velocity $u^a$. Moreover,
we have assumed quasi-neutrality, i.e., the total charge density satisfies
$\rho \approx 0$ since the electrons and ions follow the same motion. Furthermore, we
will assume that Ohm's law in the form
\begin{equation}
\label{eq:ohm}
    j^{\langle a\rangle} = \frac{1}{\mu_0\lambda}\left( E^{a} + \epsilon^{abc}v_{b}B_{c}
    \right) ,
\end{equation}
holds, where $j^{\langle a\rangle} \equiv h^{ab}j_b$ is the three-current,
$(\mu_0\lambda)^{-1}$ is the electric conductivity, $\lambda$ is the magnetic diffusivity, and $v^{a}$ is the plasma three-velocity. We note that high conductivity implies small
magnetic diffusivity. The case of non-ideal MHD in curved spacetimes has been
treated sparsely in the literature (with some exceptions \citep{fen,jed}), and the possible
effects of a finite conductivity due to gravity--electromagnetic field or fluid couplings
has therefore largely gone unnoticed.

Thus, taking the curl of the curl of $B^{a}$ and using the commutator
relation 
\begin{equation}
    \curl\,(\curl\,B)^{a} = -\D^{2}B^{a} + \D^{a}(\D_{b}B^{b}) +
    2\epsilon^{abc}\dot{B}_{\langle b\rangle}\omega_{c} +
    \mathsf{R}^{ab}B_{b} ,
    \label{eq:doublecurl1}
\end{equation}
where $\mathsf{R}_{ab}$ is defined by the relation
\begin{eqnarray}
    \mathsf{R}_{bd} &\equiv& \tfrac{2}{3}\left( \ed + \Lambda - \tfrac{1}{3}\Theta^2
     - \sigma^2 + \omega^2 \right)h_{bd} + \Pi_{bd}
     - \dot{\sigma}_{\langle bd\rangle} + \D_{\langle b}\dot{u}_{d\rangle}
    \nonumber \\ &&\,
     - \tfrac{1}{3}\Theta\left( \sigma_{bd} - \omega_{bd} \right)
     - \dot{u}_{\langle b}\dot{u}_{d\rangle}
     - 2\left( \sigma_{\langle b}\!^k\sigma_{d\rangle k}
          + \omega_{\langle b}\omega_{d\rangle}\right)
     - 2\sigma_{k[b}\omega^k\!_{d]} ,
     \label{eq:effectivecurv}
\end{eqnarray}
where $\mathscr{E} = \mathscr{E}_{\text{fluid}} + \tfrac{1}{2}(E^2 + B^2)$ is the
total energy density, and $\Pi^{ab} = \Pi^{ab}_{\text{fluid}} - (E^{\langle a}E^{b\rangle}
+ B^{\langle a}B^{b\rangle})$ is the total anisotropic pressure.
Taking the curl of Eq.\ (\ref{eq:m2}), using Eq.\ (\ref{eq:m1})
and (\ref{eq:ohm}) to replace $E^{a}$ by $B^{a}$ and $j^{\langle
a\rangle}$, we obtain
\begin{eqnarray}
    &&\!\!\!\!\!\!\!\! \curl\,(\curl\,B)^{a} =
    {\df}^{-1}\epsilon^{abc}\epsilon_{cde}\D_{b}(v^{d}B^{e}) -
    \left( \tfrac{2}{3}\Theta + {\df}^{-1} \right)\Big[
    \dot{B}^{\langle a\rangle} + \left( \tfrac{2}{3}\Theta h^{ab} -
    \sigma^{ab} + \epsilon^{abc}\omega_{c}  \right)B_{b}
    -{\df}\mu_0\epsilon^{abc}\dot{u}_{b}j_{\langle c\rangle} +
    \epsilon^{abc}\epsilon_{cde}\dot{u}_{b}v^{d}B^{e} \Big]
    \nonumber \\ &&
    - \tfrac{2}{3}\mu_0{\df}\epsilon^{abc}j_{\langle
    b\rangle}\D_{c}\Theta +
    \tfrac{2}{3}\epsilon^{abc}\epsilon_{bde}v^{d}B^{e}\D_{c}\Theta
    -
    \epsilon^{abc}\D_{b}\left[ \sigma_{cd}\left( \mu_0{\df} j^{\langle
    d\rangle} - \epsilon^{def}v_{e}B_{f} \right) \right] -
    \epsilon^{abc}\D_{b}\left[ \epsilon_{cde}\omega^{d}\left( \mu_0{\df} j^{\langle
    e\rangle} - \epsilon^{efg}v_{f}B_{g} \right)  \right]
    \nonumber \\ &&
    -
    \epsilon^{abc}\D_{b}\left(\epsilon_{cde}\dot{u}^{d}B^{e}\right)
    + {\curl}\,\dot{E}^{\langle a\rangle} .
    \label{eq:doublecurl2}
\end{eqnarray}
Furthermore, the last term in Eq.\ (\ref{eq:doublecurl2}), due to the displacement
current $\dot{E}^{\langle a\rangle}$, may be written
\begin{equation}
 {\curl}\,\dot{E}^{\langle a\rangle} =    -\ddot{B}^{\langle a\rangle}
  + \Xi^a
\end{equation}
where
\begin{eqnarray}
  && \Xi^a \equiv
  - \left( \tfrac{2}{3}\dot{\Theta}h^{ab} - \dot{\sigma}^{\langle ab\rangle}
    + \epsilon^{abc}\dot{\omega}_{\langle c\rangle} \right)B_b
  - \left( \Theta h^{ab} - \sigma^{ab}
    + \epsilon^{abc}\omega_c \right)\dot{B}_{\langle b\rangle}
  - \tfrac{1}{3}\Theta\left( \tfrac{2}{3}\Theta h^{ab} - \sigma^{ab}
    + \epsilon^{abc}\omega_c\right)B_b
 \nonumber \\ &&\quad
  - \epsilon^{abc}\dot{u}_b\left[ {\curl}\,B_c - \mu_0j_{\langle c\rangle}
    + \epsilon_{cde}\dot{u}^dB^e
    - \left( \tfrac{1}{3}\Theta h_{cd} - \sigma_{cd}
      + \epsilon_{cde}\omega^e \right)E^d \right]
  + \tfrac{1}{2}\epsilon^{abc}E_bq_c
  +\left[ 2\dot{u}^{\langle a}\omega^{b\rangle} + \D^{\langle a}\omega^{b\rangle}
   - ({\curl}\,\sigma)^{ab} \right]E_b
 \nonumber \\ &&\quad
  - \epsilon^{abc}\ddot{u}_{\langle b\rangle}E_c
  + \epsilon^{abc}\left(  \sigma_{bd} +
    \epsilon_{bde}\omega^e \right)\D^dE_c
    \label{eq:displacement}
\end{eqnarray}
by commuting spatial and time like covariant derivatives and using
Maxwell's equations. The commutation of derivatives introduces
curvature effects into the expression. Here $q^a =
q^a_{\text{fluid}} +
\epsilon^{abc}E_bB_c$ is the energy flux due to fluid and
electromagnetic (i.e., Poynting flux) contributions. The latter
derives from the gravitational self-interaction of the
electromagnetic field, and can in many cases safely be neglected.
We note that all terms in $\Xi^a$ defined by Eq.\
(\ref{eq:displacement}) are curvature contributions. The
displacement current is normally neglected in the MHD
approximation, since it corresponds to high frequency phenomena.
Here we see that the presence of gravity alters this
interpretation of the displacement current. We will be interested
in low frequency phenomena, and we therefore assume
${\curl}\,\dot{E}^{\langle a\rangle} \approx \Xi^a$.

Thus, equating (\ref{eq:doublecurl1}) with the expression
(\ref{eq:doublecurl2}), we obtain the general relativistic dynamo
equation (see also \citet{tsa2} for the source free general relativistic
electromagnetic wave equations)
\begin{eqnarray}
    && \dot{B}^{\langle a\rangle}
    - \epsilon^{abc}\epsilon_{cde}\D_{b}(v^{d}B^{e})
    - {\df}\D^{2}B^{a}
    = -\tfrac{2}{3}{\df}\Theta\dot{B}^{\langle a\rangle}  +
    2{\df}\epsilon^{abc}\omega_{b}\dot{B}_{\langle c\rangle}
    + 2{\df}\D^{a}\left[\omega_{b}\left(
    \mu_0{\df} j^{\langle b\rangle} - \epsilon^{bcd}v_{c}B_{d}\right)\right]  -
    {\df}\mathsf{R}^{ba}B_{b}
    \nonumber \\ && \quad
    -
        \left( 1 + \tfrac{2}{3}{\df}\Theta \right)\Big[
         \left( \tfrac{2}{3}\Theta h^{ab} -
        \sigma^{ab} + \epsilon^{abc}\omega_{c}  \right)B_{b}
        - \mu_0{\df}\epsilon^{abc}\dot{u}_{b}j_{\langle c\rangle} +
        \epsilon^{abc}\epsilon_{cde}\dot{u}_{b}v^{d}B^{e} \Big]
        \nonumber \\ && \quad
        -
        \tfrac{2}{3}\mu_0{\df}^{2}\epsilon^{abc}j_{\langle
        b\rangle}\D_{c}\Theta +
        \tfrac{2}{3}{\df}\epsilon^{abc}\epsilon_{bde}v^{d}B^{e}\D_{c}\Theta
        -
   {\df}\epsilon^{abc}\D_{b}\left[ \sigma_{cd}\left( \mu_0{\df} j^{\langle
        d\rangle} - \epsilon^{def}v_{e}B_{f} \right) \right]
        \nonumber \\ &&\quad
        -
   {\df}\epsilon^{abc}\D_{b}\left[ \epsilon_{cde}\omega^{d}\left( \mu_0{\df} j^{\langle
        e\rangle} - \epsilon^{efg}v_{f}B_{g} \right)  \right]
        -
   {\df}\epsilon^{abc}\D_{b}\left(\epsilon_{cde}\dot{u}^{d}B^{e}\right)
   + \lambda\Xi^a
   \label{eq:dynamo-covariant}
\end{eqnarray}
where we have used Maxwell's equation (\ref{eq:m4}) and Ohm's law
(\ref{eq:ohm}). The terms on the right hand of the dynamo equation
(\ref{eq:dynamo-covariant}) represents the influence of general
relativistic gravity, while the left hand side gives the normal
dynamo action from the fluid vorticity $\epsilon_{abc}\D^bv^c$.

Sometimes it may be advantageous to keep the electric field in the
gravitational right hand side of Eq.\ (\ref{eq:dynamo-covariant}),
in which case we have
\begin{eqnarray}
   && \dot{B}^{\langle a\rangle}
    - \epsilon^{abc}\epsilon_{cde}\D_{b}(v^{d}B^{e})
    - {\df}\D^{2}B^{a}  = -\tfrac{2}{3}{\df}\Theta\dot{B}^{\langle a\rangle}  +
    2{\df}\epsilon^{abc}\omega_{b}\dot{B}_{\langle c\rangle}
    + 2{\df}\D^{a}\left(\omega_{b}E^b \right)  -
    {\df}\mathsf{R}^{ba}B_{b}
    \nonumber \\ && \quad
    -
        \left( 1 + \tfrac{2}{3}{\df}\Theta \right)\Big[
         \left( \tfrac{2}{3}\Theta h^{ab} -
        \sigma^{ab} + \epsilon^{abc}\omega_{c}  \right)B_{b}
        - \epsilon^{abc}\dot{u}_{b}E_{c} \Big]
        -
        \tfrac{2}{3}\epsilon^{abc}E_{b}\D_{c}\Theta -
   {\df}\epsilon^{abc}\D_{b}\left( \sigma_{cd}E^d \right)
        \nonumber \\ && \quad
        -
   {\df}\epsilon^{abc}\D_{b}\left( \epsilon_{cde}\omega^{d}E^e  \right)
        -
   {\df}\epsilon^{abc}\D_{b}\left(\epsilon_{cde}\dot{u}^{d}B^{e}\right)
   +  \lambda\Xi^a.
   \label{eq:dynamo-covariant2}
\end{eqnarray}

\subsection{Three-dimensional vector notation}

The general relativistic dynamo equation may also be formulated in
terms of three-dimensional vector notation. For every spacelike
vector $X^{a}$ or PSTF tensor $Y^{ab}$ we denote the corresponding
three-dimensional vector or tensor according to $\vec{X}$ and
$\Bar{\Bar{Y}}$. With these definitions, Eq.\
(\ref{eq:dynamo-covariant2}) takes the form
\begin{eqnarray}
    && \dot{\vec{B}}
       - \vec{\nabla}\times(\vec{v}\times\vec{B})
       - {\df}\nabla^{2}\vec{B}
       = -\tfrac{2}{3}{\df}\Theta\dot{\vec{B}}  +
       2{\df}\,\vec{\omega}\times\dot{\vec{B}}
       + 2{\df}\vec{\nabla}(\vec{\omega}\cdot\vec{E})  -
       {\df}\vec{B}\cdot\Bar{\Bar{\mathsf{R}}}
       -
           \left( 1 + \tfrac{2}{3}{\df}\Theta \right)\Big(
        \tfrac{2}{3}\Theta\vec{B} -
           \Bar{\Bar{\sigma}}\cdot\vec{B} + \vec{B}\times\vec{\omega}
           - \vec{a}\times\vec{E} \Big)
       \nonumber \\ && \quad
           -
           \tfrac{2}{3}{\df}\vec{E}\times\vec{\nabla}\Theta
           -
      {\df}\vec{\nabla}\times(\Bar{\Bar{\sigma}}\cdot\vec{E})
           -
      {\df}\vec{\nabla}\times( \vec{\omega}\times\vec{E}) -
      {\df}\vec{\nabla}\times(\vec{a}\times\vec{B})
   +  \lambda\vec{\Xi} ,
      \label{eq:dynamo-3d}
\end{eqnarray}
where we have introduced the three-dimensional acceleration vector
$\vec{a}$ corresponding to $\dot{u}^{a}$, and $\vec{\nabla}$
corresponds to $\D^{a}$. Dot and cross products are defined in the
obvious way. Furthermore, the terms due to the displacement current
takes the form
\begin{eqnarray}
 &&  \vec{\Xi} =
  - \left( \tfrac{2}{3}\dot{\Theta}\vec{B} - \dot{\Bar{\Bar{\sigma}}}\cdot\vec{B}
    + \vec{B}\times\dot{\vec{\omega}} \right)
  - \left( \tfrac{2}{3}\Theta\dot{\vec{B}}  - \Bar{\Bar{\sigma}}\cdot\dot{\vec{B}}
    + \dot{\vec{B}}\times\vec{\omega}\right)
  - \dot{\vec{a}}\times\vec{E}
    +  \tfrac{1}{3}\Theta \vec{\nabla}\times\vec{E}
  + (\Bar{\Bar{\sigma}}\cdot\vec{\nabla})\times\vec{E}
  - (\vec{\omega}\times\vec{\nabla})\times\vec{E}
  \nonumber \\ &&\quad
  - \vec{a}\times\left[ \vec{\nabla}\times\vec{B} - \mu_0\vec{j}
    + \vec{a}\times\vec{B}
    - \left( \tfrac{2}{3}\Theta\vec{E} - \Bar{\Bar{\sigma}}\cdot\vec{E}
      + \vec{E}\times\vec{\omega} \right)\right]
  + \tfrac{1}{2}\vec{E}\times\vec{q}
  -\Bar{\Bar{H}}\cdot\vec{E} ,
    \label{eq:displacement2}
\end{eqnarray}
where we have introduced the magnetic part of the Weyl tensor
$H^{ab} = - 2\dot{u}^{\langle a}\omega^{b\rangle} - \D^{\langle a}\omega^{b\rangle}
   + ({\curl}\,\sigma)^{ab} $, which signifies the presence of gravitational waves or frame dragging effects.

\subsection{The equations of motion}

The dynamo equation contains the centre of mass fluid  velocity
$v^a \equiv (\ed_{(e)}v^a_{(e)} + \ed_{(i)}v^a_{(i)} )/(\ed_{(e)}
+ \ed_{(i)})$ and the current $j^{\langle a\rangle} \equiv
\rho_{(e)}v^a_{(e)} + \rho_{(i)}v^a_{(i)}$, $e$ ($i$) denoting the
electrons (ions). Using quasi-neutrality, the evolution of the
total fluid energy density (dropping the index denoting the fluid) $\ed \equiv \ed_{(e)} + \ed_{(i)}$ and the
centre of mass velocity is given by
\begin{equation}\label{eq:density}
  \dot{\ed} + \D_a(\ed v^a) = -\left(\Theta + \dot{u}_av^a\right)\ed ,
\end{equation}
and
\begin{equation}\label{eq:momentum}
  \ed\left( \dot{v}^{\langle a\rangle} + v^b\D_b v^a \right) = -\left( \tfrac{1}{3}\Theta v^a + \dot{u}^a + \sigma^a\!_b v^b + \epsilon^{abc}\omega_b v_c\right)\ed + \epsilon^{abc}j_b B_c ,
\end{equation}
respectively, in the cold plasma limit.  Moreover, the current $j^a$ can be expressed in terms of the magnetic field via Maxwell's equation (\ref{eq:m2}) and Ohm's law (\ref{eq:ohm}).

We note that the Eqs.\ (\ref{eq:density}) and (\ref{eq:momentum}) can easily be
generalised to incorporate anisotropic and/or viscous effects (see, e.g., \citet{sub}).

\section{Special cases}

In order to extract some information from the rather complicated
relativistic dynamo equation, we shall present some special cases.

\subsection{Infinite conductivity}

The large bulk of literature on general relativistic MHD has treated the case of
ideal MHD, i.e., infinite conductivity. As shown below, the general dynamo equation
greatly simplifies when this assumption is made.

As $\sigma \rightarrow \infty$, the diffusion of the magnetic
field lines decreases, i.e., ${\df} \rightarrow 0$, and we are
left with
\begin{equation}
    \dot{B}^{\langle a\rangle}
    - \epsilon^{abc}\epsilon_{cde}\D_{b}(v^{d}B^{e})
    = -  \left( \tfrac{2}{3}\Theta h^{ab} -
        \sigma^{ab} + \epsilon^{abc}\omega_{c}  \right)B_{b}
        - \epsilon^{abc}\epsilon_{cde}\dot{u}_{b}v^{d}B^{e} ,
   \label{eq:infinitecond}
\end{equation}
or
\begin{equation}
    \dot{\vec{B}} - \vec{\nabla}\times(\vec{v}\times\vec{B})
       =   - \Big[
        \tfrac{2}{3}\Theta\vec{B} -
           \Bar{\Bar{\sigma}}\cdot\vec{B} + \vec{B}\times\vec{\omega}
           + \vec{a}\times \left( \vec{v}\times\vec{B}\right) \Big]   ,
      \label{eq:infinitecond3d}
\end{equation}
which of course also can be found directly from Maxwell's
equations and the vanishing of the Lorentz force. Infinite
conductivity is a common assumption in cosmology, for example.

\subsection{Effects of expansion/collapse}

As in the case of infinite conductivity, the case of curved spacetime MHD
has largely resorted to analysing the effects of expansion/collapse, at
which the gravitational--electromagnetic field coupling is kept at a minimum.
It also gives a good description of some fundamental aspects of
cosmological MHD phenomena.

The simplest case where the nontrivial effects of expansion occur
is in homogeneous and isotropic fluid background. The homogeneity
and isotropy significantly simplifies the Einstein equations, and
the spacetime is described in terms of the scalar quantities
$\ed$, $p$, $\Theta$, and $\Lambda$. Therefore,  equations
(\ref{eq:dynamo-covariant}) and (\ref{eq:dynamo-3d}) simplify to
\begin{equation}
    \dot{B}^a
    - \epsilon^{abc}\epsilon_{cde}\D_{b}(v^{d}B^{e})
    - {\df}\D^{2}B^{a}
    = -\tfrac{2}{3}{\df}\Theta\dot{B}^a -
    {\df}\mathsf{R}^{ba}B_{b}
    -  \left( 1 + \tfrac{2}{3}{\df}\Theta \right)\tfrac{2}{3}\Theta B^{a}
    + \lambda\Xi^a,
   \label{eq:flrw-covariant}
\end{equation}
and
\begin{equation}
    \dot{\vec{B}}
       - \vec{\nabla}\times(\vec{v}\times\vec{B})
       - {\df}\nabla^{2}\vec{B}
       = -\tfrac{2}{3}{\df}\Theta\dot{\vec{B}}   -
       {\df}\vec{B}\cdot\Bar{\Bar{\mathsf{R}}}
       - \left( 1 + \tfrac{2}{3}{\df}\Theta
           \right)\tfrac{2}{3}\Theta\vec{B}
     + \lambda\vec{\Xi},
      \label{eq:flrw-3d}
\end{equation}
respectively, where $\mathsf{R}_{ab} = \tfrac{2}{3}(\mathscr{E} + \Lambda - \tfrac{1}{3}\Theta^2)h_{ab}$ is the Ricci three-curvature.
The effects of the displacement current becomes
\begin{equation}\label{eq:xi-exp-cov}
  \Xi^a = - \tfrac{2}{3}\dot{\Theta}B^a
  - \Theta\dot{B}^a - \tfrac{2}{9}\Theta^2B^a ,
\end{equation}
i.e.,
\begin{equation}\label{eq:xi-exp-3d}
  \vec{\Xi} = - \tfrac{2}{3}\dot{\Theta}\vec{B}
  - \Theta\dot{\vec{B}} - \tfrac{2}{9}\Theta^2\vec{B}
\end{equation}
Here we have neglected the gravitational self-interaction of the electromagnetic field.
Thus, from Eq.\ (\ref{eq:flrw-covariant}) and
(\ref{eq:flrw-3d}), together with (\ref{eq:xi-exp-cov}) and (\ref{eq:xi-exp-3d}), we obtain
\begin{equation}
    \dot{B}^a
    - \epsilon^{abc}\epsilon_{cde}\D_{b}(v^{d}B^{e})
    - {\df}\D^{2}B^{a}
    = -\tfrac{1}{3}\Theta(2B^a + 5\lambda\dot{B}^a)
    -\tfrac{1}{3}\lambda\left( \mathscr{E} - 3p + 4\Lambda + \tfrac{2}{3}\Theta^2 \right)B^a,
\end{equation}
and
\begin{equation}
    \dot{\vec{B}}
       - \vec{\nabla}\times(\vec{v}\times\vec{B})
       - {\df}\nabla^{2}\vec{B}
       =   -\tfrac{1}{3}\Theta(2\vec{B} + 5\lambda\dot{\vec{B}})
    -\tfrac{1}{3}\lambda\left( \mathscr{E} - 3p + 4\Lambda + \tfrac{2}{3}\Theta^2 \right)\vec{B},
\end{equation}
respectively. Thus, we see that for collapsing solutions ($\Theta < 0$), there will be a growing magnetic field.

\subsection{The effects of rotation}

In a rotating space time, i.e., $\omega^{a} \neq 0$, Einstein's
equations shows that the vorticity has as its source the
acceleration $\dot{u}^{a}$. G\"odel's universe does not require
this, as it is spacetime homogeneous, but for the sake of
completeness we will keep both spacetime acceleration and vorticity, while
neglecting the expansion and shear (thus only considering only
rigid rotation), in the equations presented below. They read
\begin{eqnarray}
    && \dot{B}^{\langle a\rangle}
    - \epsilon^{abc}\epsilon_{cde}\D_{b}(v^{d}B^{e})
    - {\df}\D^{2}B^{a}
    =  2{\df}\epsilon^{abc}\omega_{b}\dot{B}_{\langle c\rangle}
    + 2{\df}\D^{a}\left[\omega_{b}\left(
    \mu_0{\df} j^{\langle b\rangle} - \epsilon^{bcd}v_{c}B_{d}\right)\right]  -
    {\df}\mathsf{R}^{ba}B_{b}
    -
   {\df}\epsilon^{abc}\D_{b}\left(\epsilon_{cde}\dot{u}^{d}B^{e}\right)
    \nonumber \\ && \quad
    - \Big[- \epsilon^{abc}\omega_{b} B_{c}
        - \mu_0{\df}\epsilon^{abc}\dot{u}_{b}j_{\langle c\rangle} +
        \epsilon^{abc}\epsilon_{cde}\dot{u}_{b}v^{d}B^{e} \Big]
        -
   {\df}\epsilon^{abc}\D_{b}\left[ \epsilon_{cde}\omega^{d}\left( \mu_0{\df} j^{\langle
        e\rangle} - \epsilon^{efg}v_{f}B_{g} \right)  \right]
          + \lambda\Xi^a
   \label{eq:rotation-covariant}
\end{eqnarray}
or
\begin{eqnarray}
    && \dot{\vec{B}}
       - \vec{\nabla}\times(\vec{v}\times\vec{B})
       - {\df}\nabla^{2}\vec{B}
       = 2{\df}\,\vec{\omega}\times\dot{\vec{B}}
       + 2{\df}\vec{\nabla}\left[\vec{\omega}\cdot\left(
       \mu_0{\df} \vec{j} - \vec{v}\times\vec{B}\right)\right]  -
       {\df}\vec{B}\cdot\Bar{\Bar{\mathsf{R}}}
       - \Big[ \vec{B}\times\vec{\omega}
           - \vec{a}\times\left(\mu_0{\df}\vec{j} -
           \vec{v}\times\vec{B}\right) \Big]
           \nonumber \\ &&\quad
           -
      {\df}\vec{\nabla}\times\left[ \vec{\omega}\times\left(
      \mu_0{\df}\vec{j} - \vec{v}\times\vec{B} \right)  \right] -
      {\df}\vec{\nabla}\times\left(\vec{a}\times\vec{B}\right)
   + \lambda\vec{\Xi} .
      \label{eq:rotation-3d}
\end{eqnarray}
We note that the contribution $\Xi^a$ from the displacement current is
only slightly simplified compared to the generic case. Thus, as in many
problems of general relativity, rotation of spacetime gives rise to highly
complex equations, and the dynamo equation is no exception.

\section{Gravitational waves}

Linear gravitational waves on a homogeneous background can be
covariantly defined as tensor perturbations satisfying
$\D_{a}\sigma^{ab} =
\D_{a}E^{ab} = \D_{a}H^{ab} = 0$, thus being transverse and
traceless. In order to clearly see the effects of gravitational
waves on the induction of electromagnetic fields, we consider
this perturbation on a Minkowski
background. In this case, the dynamo equation becomes
\begin{equation}
   \dot{B}^{\langle a\rangle}
    - \epsilon^{abc}\epsilon_{cde}\D_{b}(v^{d}B^{e})
    - {\df}\D^{2}B^{a} = -{\df}\mathsf{R}^{ba}B_{b}
    + \sigma^{ab}B_{b}
    -  {\df}\epsilon^{abc}\D_{b}( \sigma_{cd}E^d )
   + \lambda\Xi^a,
   \label{eq:gw-covariant}
\end{equation}
or
\begin{equation}
    \dot{\vec{B}}
       - \vec{\nabla}\times(\vec{v}\times\vec{B})
       - {\df}\nabla^{2}\vec{B}
       = - {\df}\vec{B}\cdot\Bar{\Bar{\mathsf{R}}}
       + \Bar{\Bar{\sigma}}\cdot\vec{B}
           -
      {\df}\vec{\nabla}\times( \Bar{\Bar{\sigma}}\cdot\vec{E} )
  + {\df}\vec{\Xi}  ,
      \label{eq:gw-3d}
\end{equation}
where we have neglected all products of gravitational variables.

The contribution from $\Xi^a$ becomes
\begin{equation}\label{eq:xi-gw}
 \Xi^a = \dot{\sigma}^{\langle ab\rangle}B_b
         + \sigma^{ab}\dot{B}_{\langle b\rangle}
         + \epsilon^{abc}\sigma_{bd}{\D}^dE_c
         - ({\curl}\,\sigma)^{ab}E_b .
\end{equation}

Equation (\ref{eq:effectivecurv}) gives an expression for the
Ricci three-curvature in terms of the shear of the perturbation.
Thus, using (\ref{eq:xi-gw}), Eqs.\ (\ref{eq:gw-covariant}) and
(\ref{eq:gw-3d}) give us
\begin{equation}
   \dot{B}^{\langle a\rangle}
    - \epsilon^{abc}\epsilon_{cde}\D_{b}(v^{d}B^{e})
    - {\df}\D^{2}B^{a} = \sigma^{ab}B_{b}
    + {\df}\left[ 2\dot{\sigma}^{\langle ab\rangle}B_{b}
    -  \epsilon^{abc}\D_{b}( \sigma_{cd}E^d)
         + \sigma^{ab}\dot{B}_{\langle b\rangle}
         + \epsilon^{abc}\sigma_{bd}\D^dE_c
         - H^{ab}E_b \right],
   \label{eq:gw-covariant2}
\end{equation}
or
\begin{equation}
    \dot{\vec{B}}
       - \vec{\nabla}\times(\vec{v}\times\vec{B})
       - {\df}\nabla^{2}\vec{B}
       =    \Bar{\Bar{\sigma}}\cdot\vec{B}
    + {\df}\left[ 2\dot{\Bar{\Bar{\sigma}}}\cdot\vec{B}
    -  \vec{\nabla}\times( \Bar{\Bar{\sigma}}\cdot\vec{E})
         + \Bar{\Bar{\sigma}}\cdot\dot{\vec{B}}
         + (\Bar{\Bar{\sigma}}\cdot\vec{\nabla})\times\vec{E}
         - \Bar{\Bar{H}}\cdot\vec{E}\right] ,
      \label{eq:gw-3d2}
\end{equation}
respectively, where $H^{ab} = (\curl\,\sigma)^{ab}$.
The terms on the rhs describe the effect of the curvature, and
given a GW they will drive the magnetic field evolution.
We see that the finite electric conductivity gives rise to new couplings
between GWs and magnetic fields.

\subsection{Gravitational waves in a constant magnetic field}

The case of a GW passing through an initially constant magnetic
field will help illustrate these equations. We treat all disturbances
$\nabla B$ in the magnetic field as well as the gravitational wave variables
as first order perturbations, and we therefore neglect all terms in
the equations which are a product of derivatives of the magnetic
field with GW terms ($\sigma_{ab}$), and terms ${\cal O}([\nabla
B]^2)$. We may also neglect disturbances in the energy density of
the fluid. With these approximations, we find that the momentum
conservation equation (\ref{eq:momentum}) becomes
\begin{equation}\label{eq:velocity}
  \ed\dot{{v}}^{a} =  \epsilon^{abc}j_b B_c ,
\end{equation}
while the current is given by Eq.\ (\ref{eq:m2}):
\begin{equation}\label{eq:current}
  {j}^{ a} = \mu_0^{-1}\,{\curl}\,{B}^{a}.
\end{equation}
The dynamo equation (\ref{eq:gw-covariant2}) simplifies to
\begin{equation}
   \dot{B}^{ a}
    - \epsilon^{abc}\epsilon_{cde}B^{e}\D_{b}v^{d}
    - {\df}\D^{2}B^{a} = 2{\df}\dot{\sigma}^{
    ab}B_{b} + \sigma^{ab}B_{b}.
   \label{eq:gw-covariant-simp}
\end{equation}
We may decouple Eqs.\ (\ref{eq:velocity})--(\ref{eq:gw-covariant-simp})
by taking the time derivative of Eq.~(\ref{eq:gw-covariant-simp}):
\begin{equation}\label{eq:dynamo-gw}
\ddot B^a-\lambda\D^2\dot B^a+\frac{1}{\mu_0\mathscr{E}}B^b\left[
-B^a\D^2B_b+B^c(\D^a\D_b
B_c-\D_b\D_cB^a)\right]
= 2{\df}\ddot{\sigma}^{
    ab}B_{b} + \dot\sigma^{ab}B_{b}
\end{equation}
where (neglecting backreaction from the electromagnetic  field)
the shear perturbation is determined by the source-free wave
equation
\begin{equation}\label{eq:gw}
  \ddot{{\sigma}}^{ab} - \D^2{\sigma}^{ab} = 0 .
\end{equation}

We split the perturbed magnetic field parallel and perpendicular
to the `background' magnetic field according to
\begin{equation}
  {B}^a = B_0\left[(1+\mathscr{B}) e^a + \mathscr{B}^a  \right] ,
  \quad \mathscr{B}^a \equiv N^{ab}{B}_b/B_0 ,
  \quad \text{and}
  \quad \mathscr{B} \equiv \left(e_a{B}^a-1\right)/B_0 .
\end{equation}
where $N_{ab} = h_{ab} - e_ae_b$ and $e^a$ is a spacelike unit
vector: $e_ae^a=1$~(see \cite{CB} for a detailed discussion of
this split). Here, $B_0$ is a constant, denoting the magnitude of
the magnetic field when the GW is zero (in which case $B^a=B_0
e^a$). The perturbations in the static magnetic field are given by
$\mathscr{B}$, parallel to the background field, and
$\mathscr{B}^a$ perpendicular to it. We neglect all products of
$\mathscr{B}$ and $\mathscr{B}^a$, and terms of the form $\mathscr{B}\sigma$ (note that
$\nabla e={\cal O}(\sigma)$), since these are of higher order.

The shear perturbation can be decomposed according to
\begin{equation}\label{eq:gw-split}
  {\sigma}_{ab} = \Sigma_{ab} + 2\Sigma_{(a}e_{b)}
  + \left( e_ae_b - \tfrac{1}{2}N_{ab} \right)\Sigma
\end{equation}
by introducing the gravitational wave variables
\begin{equation}
  \Sigma \equiv {\sigma}^{ab}e_ae_b ,
  \quad \Sigma_a \equiv N_{ab}{\sigma}^{bc}e_c ,
  \quad \text{and}
  \quad \Sigma_{ab} \equiv \left( N_{(a}\!^cN_{b)}\!^d
  - \tfrac{1}{2}N_{ab}N^{cd} \right){\sigma}_{cd},
\end{equation}
and the operators
\begin{equation}
  \quad \D_{\|} \equiv e^a\D_a =\frac{\mathrm d~}{\mathrm{d}z}
  \quad \text{and}
  \quad \D^a_{\perp} \equiv N^{ab}\D_b ,
\end{equation}
allow us to split the spatial derivative along, and perpendicular
to, the unperturbed magnetic field.\footnote{Note that the
definition of $\D_\perp^a$ for tensors has projection with
$N_{ab}$ on each free index \citep{CB}.}

Using these definitions, and the frame choice\footnote{We refer to
\citet{CB} for a full discussion of frame choice vs. true GW
degrees of freedom.} $\dot e^a=0=\D_\perp^ae_a$, we may split
Eq.~(\ref{eq:dynamo-gw}) into a longitudinal scalar equation along
$e^a$:
\begin{equation}\label{eq:scalar}
  \ddot{\mathscr{B}} - \lambda\D^2\dot{\mathscr{B}} - C_A^2\D^2\mathscr{B} =
  2\lambda\ddot{\Sigma}
  + \dot{\Sigma} ,
\end{equation}
where $C_A \equiv (\epsilon_0B_0^2/\ed )^{1/2}$ is the Alfv\'en
velocity of the MHD plasma, and a transverse vector equation
perpendicular to $e^a$:
\begin{equation}\label{eq:vector}
  \ddot{\mathscr{B}}^a - \lambda\D^2\dot{\mathscr{B}}^a - C_A^2\left(
  \D_{\|}^2\mathscr{B}^a - \D^a_{\perp}\D_{\|}\mathscr{B} \right) =
  2\lambda \ddot{\Sigma}^a + \dot{\Sigma}^a.
\end{equation}
We also have the  $\D_aB^a=0$ constraint, which takes the simple form
$\D_\|\mathscr{B}+\D_\perp^a\mathscr{B}_a=0$ using our new variables.
Neglecting
backreaction, gravitational waves propagating along the background
magnetic field will not affect the magnetic field evolution, due
to the condition of vanishing divergence $\D_a{\sigma}^{ab} = 0 =
e_a{\sigma}^{ab}$. This can also be seen from Eq.\
(\ref{eq:gw-split}), where (if $e^a$ represents the propagation
direction of the gravitational wave) $\Sigma_{ab}$ is the only
non-vanishing part of the decomposition of ${\sigma}_{ab}$, thus
removing the coupling to the magnetic field via $\Sigma$ and
$\Sigma^a$. On the other hand, including the backreaction from the
magnetic field on the gravitational wave will in general make the
gravitational wave non-transverse, by the introduction of
small scale tidal forces as well as large scale curvature effects,
but this issue is left for future research.

Next we introduce spatial harmonics for the perturbed variables
(see~\cite{CB}). Define the scalar harmonics
\begin{equation}
  D_{\perp}^2Q_{(k_\perp)} = -k_{\perp}^2Q_{(k_\perp)},
\end{equation}
where $k_\perp$ is the harmonic number for the part of the fields
perpendicular to $e^a$. We can then define vector harmonics of
even and odd parity,
\begin{equation}
  Q_{(k_\perp)}^a = (k_{\perp})^{-1}D_{\perp}^aQ_{(k_\perp)}
  \quad
  \bar{Q}_{(k_\perp)}^a = (k_{\perp})^{-1}\epsilon^a\!_bD_{\perp}^bQ_{(k_\perp)} ,
\end{equation}
respectively, where $\epsilon_{ab}=\epsilon_{abc}e^c =
-\epsilon_{ba}$ is the volume element on two-surfaces orthogonal
to $e^a$. These vector harmonics obey $D_{\perp}^2Q_{(k_\perp)}^a
= -k_{\perp}^2Q_{(k_\perp)}^a$ (similarly for
$\bar{Q}_{(k_\perp)}^a$).

Using these definitions, we may expand the magnetic field
variables as
\begin{eqnarray}\label{eq:hd-vector}
  \mathscr{B} &=& \sum_{k_\perp}\mathscr{B}_{(k_\perp)}^SQ_{(k_\perp)},\\
  \mathscr{B}^a &=& \sum_{k_\perp} \mathscr{B}_{(k_\perp)}^VQ_{(k_\perp)}^a +
  \bar{\mathscr{B}}_{(k_\perp)}^V\bar{Q}_{(k_\perp)}^a ,
\end{eqnarray}
and similarly for the GW shear variables
\begin{equation}\label{eq:hd-gw}
  \Sigma = \sum_{k_\perp} \Sigma_{(k_\perp)}^SQ_{(k_\perp)} ,
  \quad \text{and}
  \quad
  \Sigma^a = \sum_{k_\perp}  \Sigma_{(k_\perp)}^VQ_{(k_\perp)}^a
  + \bar{\Sigma}_{(k_\perp)}^V\bar{Q}_{(k_\perp)}^a  .
\end{equation}
Using these harmonic expansions,  Eq.~(\ref{eq:scalar})  can now
be written as a partial differential equation for each $k_\perp$,
using $'=\partial_z$:
\begin{equation}
\ddot{\mathscr{B}}_S - C_A^2\mathscr{B}_S'' - \lambda \dot{\mathscr{B}}_S'' +
k_\perp^2\lambda\dot{\mathscr{B}}_S + k_\perp^2C_A^2
\mathscr{B}_S = 2\lambda\ddot\Sigma_S + \dot\Sigma_S, \label{eq:scalar-wave}
\end{equation}
while the even part of (\ref{eq:vector}) may be derived from
$k_\perp\mathscr{B}_V=\mathscr{B}_S'$ (the div-B constraint), and
is in fact the same equation, but with $S$ replaced by $V$; these
give the dynamics of the `even sector', while for the odd sector
we have
\begin{equation}\label{eq:odd-vector-wave}
\ddot{\bar{\mathscr{B}}}_V-C_A^2{\bar{\mathscr{B}}}_V''-\lambda \dot{\bar{\mathscr{B}}}_V''
+ k_\perp^2\lambda\dot{\bar{\mathscr{B}}}_V + k_\perp^2 C_A^2
{\bar{\mathscr{B}}}_V= 2\lambda\ddot{\bar\Sigma}_V + \dot{\bar
\Sigma}_V.
\end{equation}
The GW variables obey
$\ddot\Sigma-D_\|^2\Sigma_S+k_\perp^2\Sigma_S=0$, and similarly
for $\Sigma_V$ and $\bar\Sigma_V$; $\D^b\sigma_{ab}=0$ implies
$k_\perp\Sigma_V=\D_\|\Sigma_S$. We have dropped $k_\perp$
subscripts, with the understanding that these equations apply for
each value of $k_\perp$. We may now introduce harmonics for
$\partial_z$ and $\partial_t$ in the usual way if desired.

\subsection{Integration}

As a means of understanding the effect of conductivity on a
magnetic field in a GW spacetime we shall consider the case of a
Gaussian pulse of GW travelling in the positive $z$-direction
through a static magnetic field such that at $t=0$,
\begin{eqnarray}
{\Sigma}^{(k)}_V&=&\Sigma e^{-z^2/L^2}\cos\left(\omega z\right),\\
\dot{\Sigma}^{(k)}_V&=&\Sigma e^{-z^2/L^2}\left[\omega\sin\left(\omega z\right)
+2\frac{z}{L}\cos\left(\omega z\right)\right],
\end{eqnarray}
where we set $\Sigma=-1$ without loss of generality
(alternatively, this could be $\bar\Sigma_V$ because the decoupled
equation for $\bar{\mathscr{B}}_V$ is identical to that of
$\mathscr{B}_V$). This means that $\mathscr{B}$ is measured in
units of the GW amplitude in the plots below. We will now
integrate Eq:~(\ref{eq:scalar-wave}) for a variety of situations
assuming that at $t=0$, $\mathscr{B}_S=\mathscr{B}_V=0$~-- i.e.,
the interaction switches on at $t=0$. We take $\omega=\pi/3$,
$k_\perp=0$ (so $\mathscr{B}_S=0$) and $L=8$, below, all in units
where $c=1$.

\begin{figure}
\[
\begin{array}{lr}
\includegraphics[width=.45\textwidth]{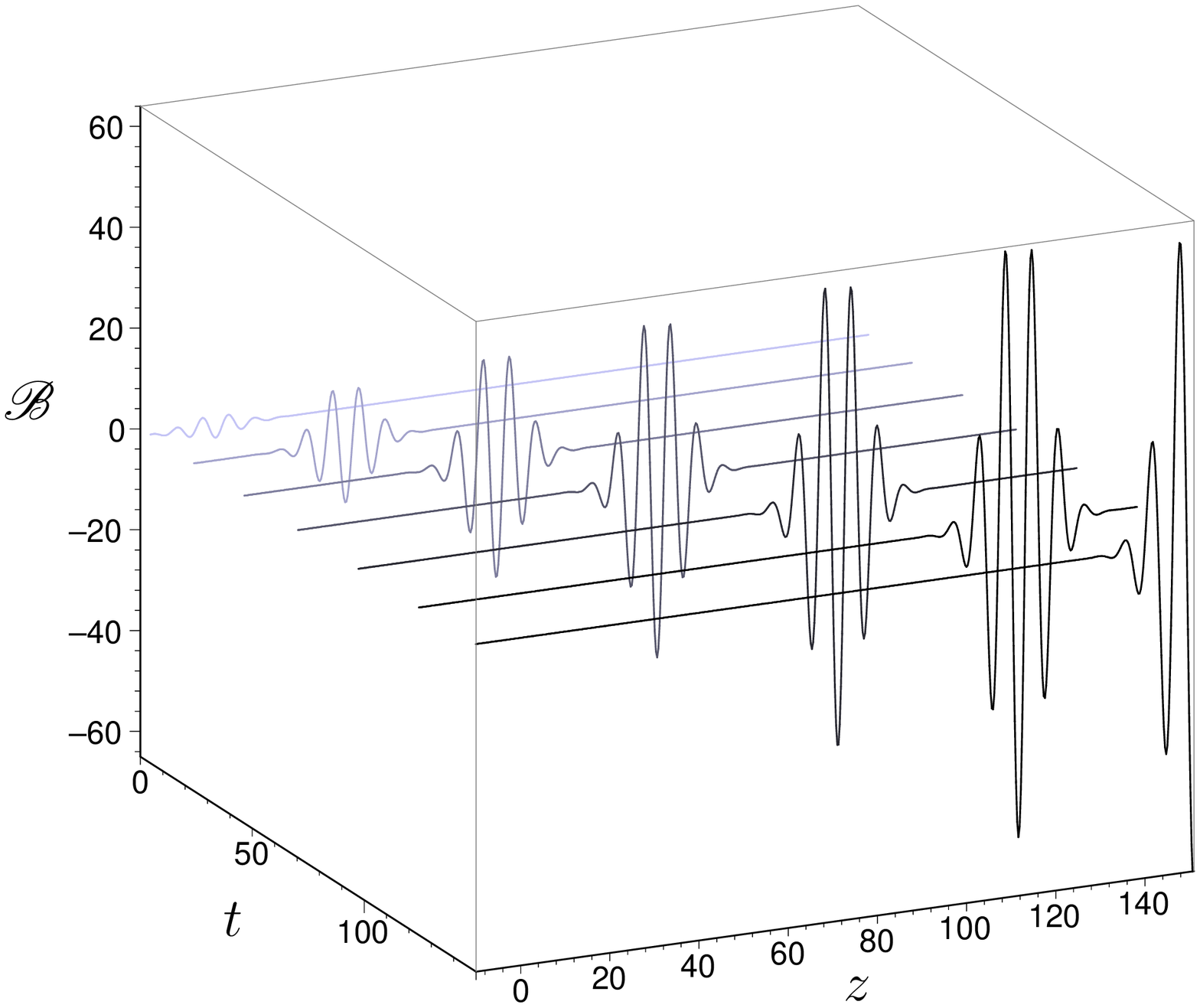}\hspace{5mm}
&
\hspace{5mm}\includegraphics[width=.45\textwidth]{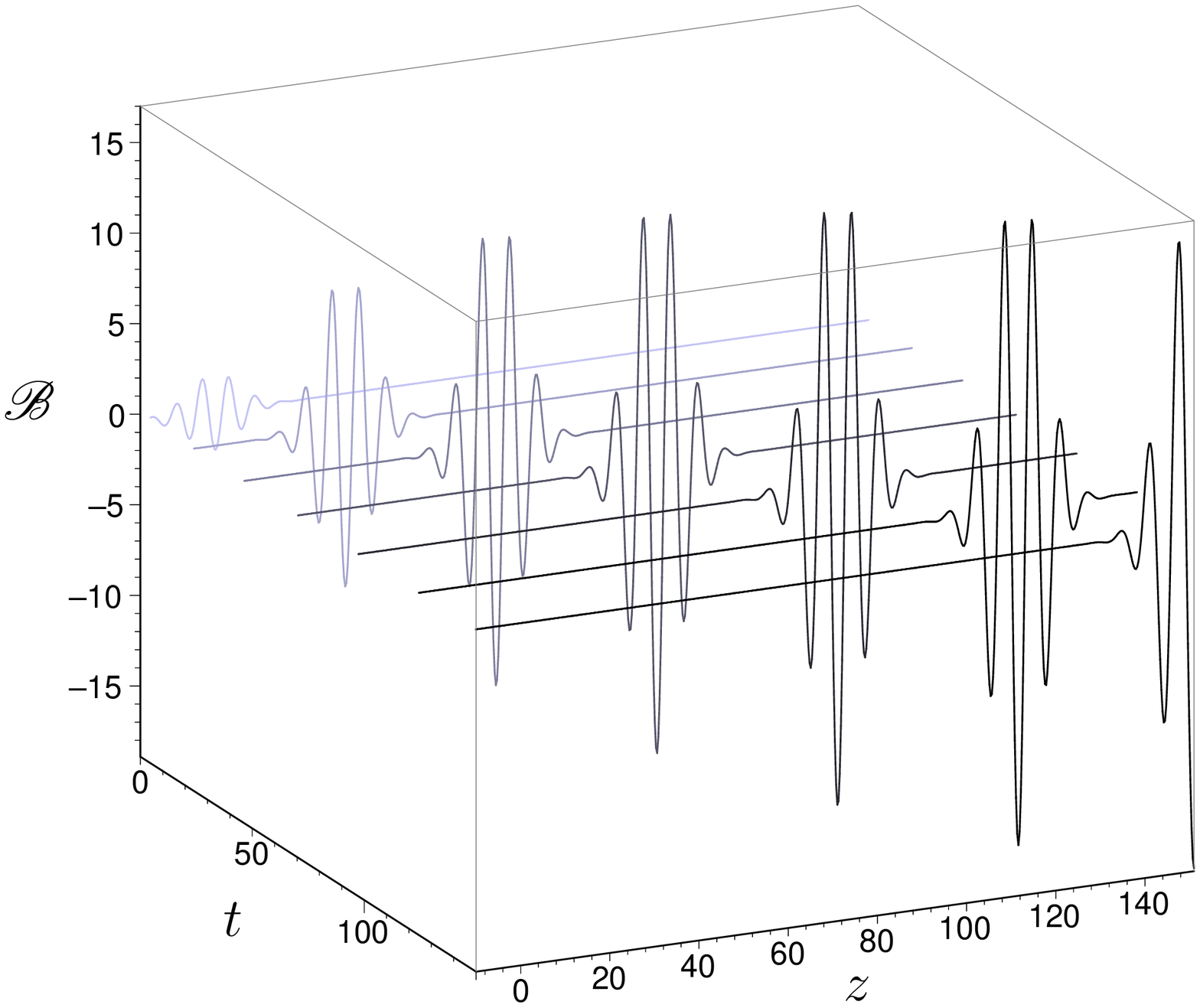}
\end{array}
\]
\caption{\small The effect of $\lambda$ on long wavelength modes, with $C_A=1$.
The plot on the left has $\lambda=0$ showing linear growth of the
magnetic field as the GW passes through the background field. On
the right, we have the same situation but with finite
conductivity, $\lambda=0.05$. While there is some early
amplification of the magnetic field, this quickly saturates, with
late time behaviour just a brief disturbance as the GW passes.
Note the difference of the vertical scale between the two
plots.\label{fig:long-wavelength} }
\end{figure}

\begin{figure}
\[
\begin{array}{lr}
\includegraphics[width=.45\textwidth]{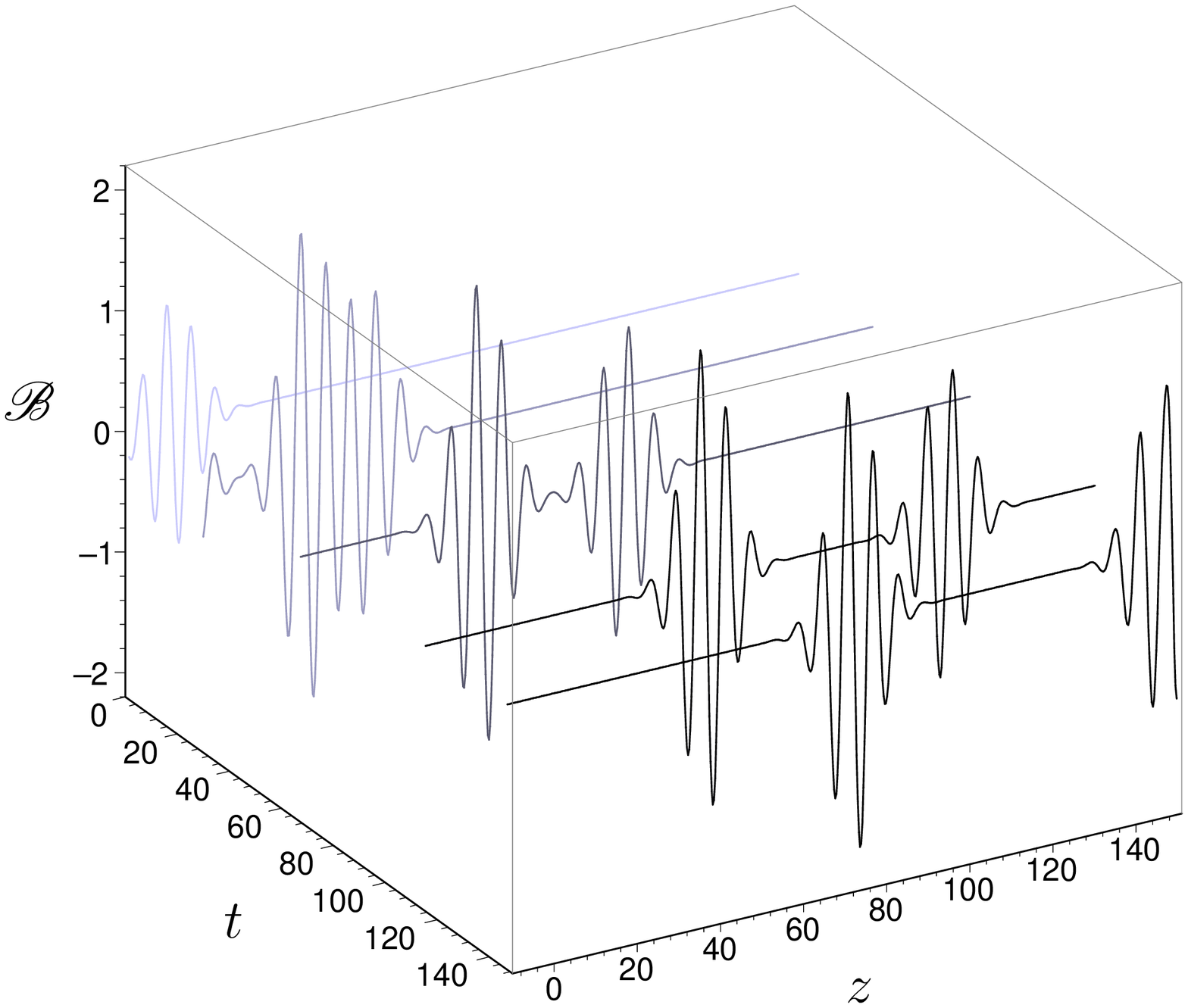}\hspace{5mm}
&
\hspace{5mm}\includegraphics[width=.45\textwidth]{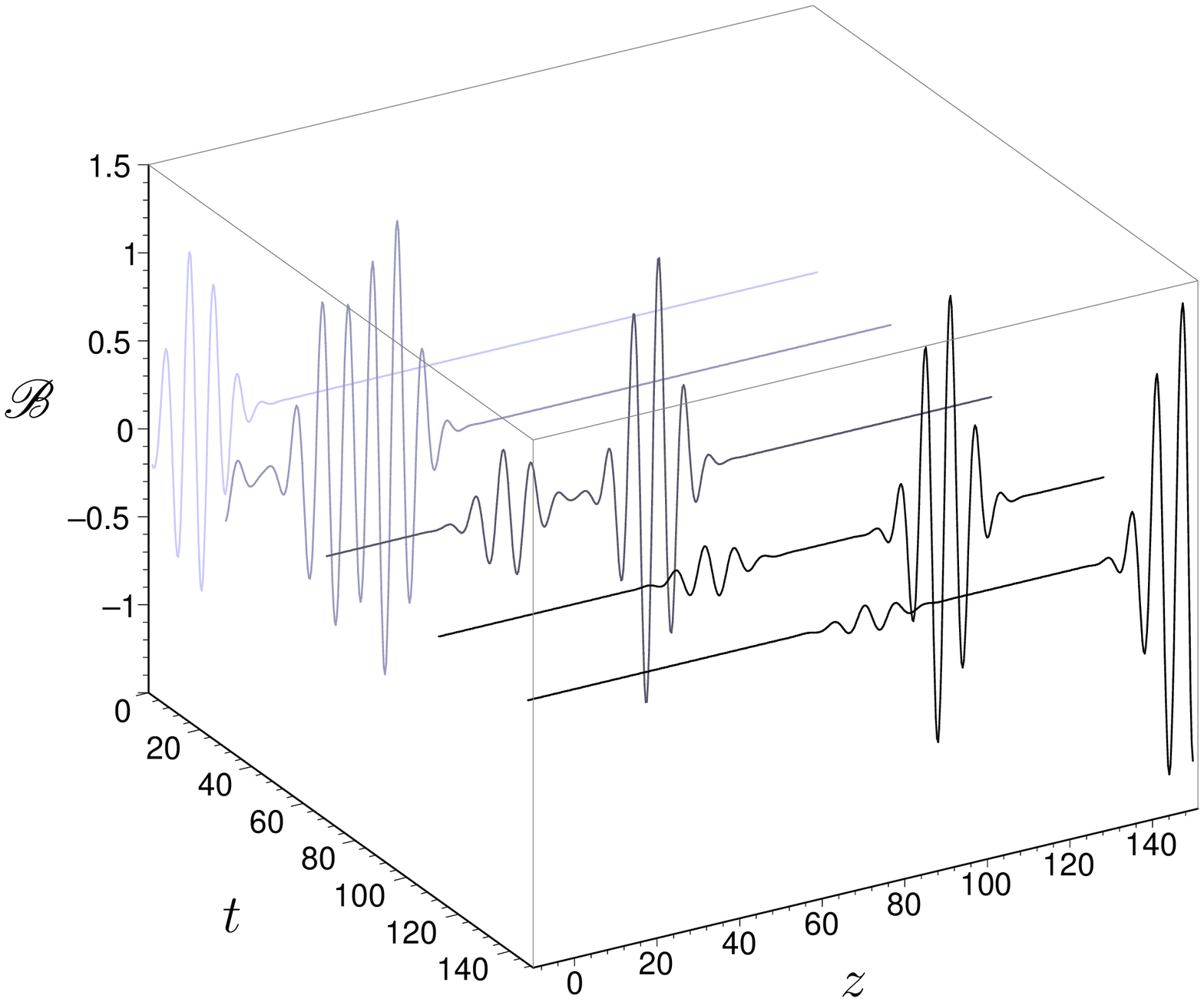}
\end{array}
\]
\caption{\small The effect of $\lambda$ on slow Alfv\'en waves, with $C_A=0.5$.
The plot on the left has $\lambda=0$ showing little growth of the
magnetic field as the GW passes through the background field,
trailed by a slow wave. The dispersion takes the form of a slower
wave trailing the oscillations induced by the GW. On the right, we
have the same scenario but with finite conductivity,
$\lambda=0.05$. The trailing Alfv\'en waves are dissipated very
quickly. The main pulse, as in the previous plot, is from a direct
forcing from the GW. \label{fig:short-wavelength}}
\end{figure}

The qualitative features of finite conductivity are shown in
figures~\ref{fig:long-wavelength} and ~\ref{fig:short-wavelength}.

In Fig.~\ref{fig:long-wavelength} we consider the case of $C_A=1$,
that is waves in a very tenuous plasma, giving rise to resonant
magnetic field amplification. The figures shown apply equally
to $\mathscr{B}_V$ or $\bar{\mathscr{B}}_V$. In the case of
infinite conductivity, shown on the left, we have a linearly
growing magnetic field pulse, which is continually sourced by the
GW as it travels through the background field. When conductivity
is present, however, this amplification is inhibited in that the
induced magnetic field can only grow to a certain strength
(compare the scales of the two plots) before a saturation
amplitude is met, at an amplitude considerably less than if the
magnetic diffusivity is zero.

If the magnetic field is strong, or the fluid dense, then the
Alfv\'en velocity can be small, resulting in weak non-resonant
amplification of GW.  The plot on the left of
Fig.~\ref{fig:short-wavelength} shows the modes with $C_A=0.5$
when $\lambda=0$. Dispersion shows up here as a trail of small
amplitude waves, trailing behind the initial pulse, due to the
mismatch in the GW and plasma dispersion relations. The plot on
the right demonstrates what happens when the conductivity is
finite. As in the resonant case the forward driven pulse is damped
and can only reach a certain amplitude, which is considerably less
than when $C_A=1$. The trailing part of the wave has nothing to
sustain it, so diffuses and rapidly decays away.

\section{Discussion}

Finite conductivity can alter the behaviour of conducting fluids,
plasmas and electromagnetic fields in a vast variety of ways. When
spacetime curvature may be neglected these effects are well
understood in a significant number of physical situations, ranging
from the early universe, to astrophysical and laboratory plasmas.
When spacetime curvature effects are present and significant,
however, very little is known about the role played by finite
conductivity in astrophysics and cosmology. We present here, for
the first time, the dynamo equation in its full generality in an arbitrary
curved spacetime.

The MHD approximation, describing low frequency charged fluid
phenomena,  allows us to neglect the displacement current in
Amp\`ere's Law, from which we may derive a diffusion equation for
the magnetic field~-- the dynamo
equation~(\ref{eq:dynamo-covariant})
or~(\ref{eq:dynamo-covariant2}). For easy comparison with
non-general relativistic results we have utilised the intuitive
1+3~covariant approach to relativistic analysis~\citep{EvE}. This
approach splits spacetime into space and time in a covariant way
by use of a timelike vector field, which then also plays the role
of the convective derivative, but now on an arbitrary curved
manifold. Spacetime dynamics is then covariantly described by the
use of invariantly defined scalars, 3-vectors, and PSTF tensors,
made up from the Riemann curvature tensor, and the dynamical
quantities associated with $u^a$. These quantities feed into
Maxwell's equations in a non-trivial way, making the resultant
electromagnetic field experience `gravitational currents' of a
very different nature from flat space physics.

In the relativistic dynamo equation we have derived,  these
gravitational currents come in a huge variety of terms which are
very difficult to analyse in general. We have presented in some
simplifying assumptions which are often used in spacetime
modelling, in order to better gain an intuition about the kinds of
effects conductivity coupled to gravity can produce.

As a specific example of analysing the relativistic dynamo
equation, we considered in detail the case of a plane GW
interacting with a static magnetic field. We believe this may play
an significant role around compact objects, where magnetic fields
and gravity waves can interact strongly. By introducing covariant
vector harmonics adapted to the magnetic field we reduced the
dynamo equation to two fairly simple coupled PDEs with a
constraint, and one decoupled PDE, with the space dimension along
the background field. These PDEs for the induced magnetic field
are sourced by the GW in an intuitive fashion, and may be analysed
by standard means. We integrated these equations numerically for a
Gaussian pulse of radiation passing through the background static
magnetic field; the summary of this integration is presented in
the figures. In the case of infinite conductivity, there is linear
amplification of the magnetic field which mimics the GW; this is
followed by a trail of oscillations resulting from waves in the
magnetic field travelling with the Alfv\'en speed. When magnetic
diffusivity is present, these trailing modes diffuse quickly, and
the overall amplification of the magnetic field is inhibited; a
ripple in the magnetic field is forced as the GW passes through.

\end{document}